\documentclass[conference]{IEEEtran}
\ifCLASSINFOpdf
   \usepackage[pdftex]{graphicx}
   \graphicspath{{./}}
   \DeclareGraphicsExtensions{.pdf,.jpeg,.png}
\else
  \usepackage[dvips]{graphicx}
  \graphicspath{{./}}
  \DeclareGraphicsExtensions{.ps}
\fi
\usepackage{url}
\usepackage[utf8]{inputenc}
\usepackage[capitalise]{cleveref}

\usepackage{color}
\usepackage{xargs}

\usepackage[colorinlistoftodos,prependcaption,textsize=tiny]{todonotes}
\newcommandx{\unsure}[2][1=]{\todo[linecolor=red,backgroundcolor=red!25,bordercolor=red,#1]{#2}}
\newcommandx{\change}[2][1=]{\todo[linecolor=blue,backgroundcolor=blue!25,bordercolor=blue,#1]{#2}}
\newcommandx{\info}[2][1=]{\todo[linecolor=yellow,backgroundcolor=yellow!25,bordercolor=yellow,#1]{#2}}
\newcommandx{\improvement}[2][1=]{\todo[linecolor=green,backgroundcolor=green!25,bordercolor=green,#1]{#2}}
\newcommandx{\thiswillnotshow}[2][1=]{\todo[disable,#1]{#2}}

\begin{document}
%
% paper title
% Titles are generally capitalized except for words such as a, an, and, as,
% at, but, by, for, in, nor, of, on, or, the, to and up, which are usually
% not capitalized unless they are the first or last word of the title.
% Linebreaks \\ can be used within to get better formatting as desired.
% Do not put math or special symbols in the title.
\title{Distributed Ledger Technology: Blockchain Compared to Directed Acyclic Graph}

% author names and affiliations
% use a multiple column layout for up to three different
% affiliations
\author{\IEEEauthorblockN{Federico Matteo Benčić and Ivana Podnar Žarko}
\IEEEauthorblockA{University of Zagreb, Faculty of Electrical Engineering and Computing
\\
Email: federico-matteo.bencic@fer.hr, ivana.podnar@fer.hr}
}

% conference papers do not typically use \thanks and this command
% is locked out in conference mode. If really needed, such as for
% the acknowledgment of grants, issue a \IEEEoverridecommandlockouts
% after \documentclass

% for over three affiliations, or if they all won't fit within the width
% of the page, use this alternative format:
% 
%\author{\IEEEauthorblockN{Michael Shell\IEEEauthorrefmark{1},
%Homer Simpson\IEEEauthorrefmark{2},
%James Kirk\IEEEauthorrefmark{3}, 
%Montgomery Scott\IEEEauthorrefmark{3} and
%Eldon Tyrell\IEEEauthorrefmark{4}}
%\IEEEauthorblockA{\IEEEauthorrefmark{1}School of Electrical and Computer Engineering\\
%Georgia Institute of Technology,
%Atlanta, Georgia 30332--0250\\ Email: see http://www.michaelshell.org/contact.html}
%\IEEEauthorblockA{\IEEEauthorrefmark{2}Twentieth Century Fox, Springfield, USA\\
%Email: homer@thesimpsons.com}
%\IEEEauthorblockA{\IEEEauthorrefmark{3}Starfleet Academy, San Francisco, California 96678-2391\\
%Telephone: (800) 555--1212, Fax: (888) 555--1212}
%\IEEEauthorblockA{\IEEEauthorrefmark{4}Tyrell Inc., 123 Replicant Street, Los Angeles, California 90210--4321}}

% use for special paper notices
%\IEEEspecialpapernotice{(Invited Paper)}

% make the title area
\maketitle

% As a general rule, do not put math, special symbols or citations
% in the abstract
\begin{abstract}
Nowadays, blockchain is becoming a synonym for distributed ledger technology. However, blockchain is only one of the specializations in the field and is currently well-covered in existing literature, but mostly from a cryptographic point of view. Besides blockchain technology, a new paradigm is gaining momentum: directed acyclic graphs. The contribution presented in this paper is twofold. Firstly, the paper analyzes distributed ledger technology with an emphasis on the features relevant to  distributed systems. Secondly, the paper analyses the usage of directed acyclic graph paradigm in the context of distributed ledgers, and compares it with the blockchain-based solutions. The two paradigms are compared using representative implementations: Bitcoin, Ethereum and Nano. We examine representative solutions in terms of the applied data structures for maintaining the ledger, consensus mechanisms, transaction confirmation confidence, ledger size, and scalability.

\end{abstract}

% no keywords
% distributed ledger, scalability, Bitcoin, Ethereum, Nano 

% For peer review papers, you can put extra information on the cover
% page as needed:
% \ifCLASSOPTIONpeerreview
% \begin{center} \bfseries EDICS Category: 3-BBND \end{center}
% \fi
%
% For peerreview papers, this IEEEtran command inserts a page break and
% creates the second title. It will be ignored for other modes.
\IEEEpeerreviewmaketitle

\section{Introduction}
% no \IEEEPARstart

Distributed ledger technology (DLT) enables the maintenance of a global, append only, data structure by a set of mutually untrusted participants in a distributed environment~\cite{Narayanan2017}. The most notable features of distributed ledgers are immutability, resistance to censorship, decentralized maintenance, and elimination of the need for a centralized trusted third party. In other words, there is no need for an entity to be in charge of conflict resolution and upkeep of a global truth that is trusted by all stakeholders which do not trust each other. Distributed ledger is suitable for tracking the ownership of digital assets, and hence it's most prominent application is the Bitcoin network~\cite{Franco2014}. DLT holds promise beyond mere cryptocurrency transfer since an entry in the ledger may be generalized to hold arbitrary data. However, before being applicable on a global scale, DLT needs to solve a number of issues it is currently facing. Blockchains, a specialization of DLTs, are getting a new rival in the field: distributed acyclic graphs (DAG). The most notable difference between the two is that blockchains bundle transactions in cryptographically linked blocks forming a single chain containing the global truth, while DAGs use a graph where a transaction is represented as a node in the graph.

This paper compares the two DLT paradigms by focusing on features relevant to their distributed design, and explains how the two tackle some of the known issues distributed ledgers are facing. In particular, we examine the applied data structures for ledger maintenance, consensus mechanisms, transaction confirmation confidence, as well as ledger size and scalability issues. A comparative qualitative analysis is presented using three reference implementations: Bitcoin~\cite{Franco2014} and Ethereum~\cite{Buterin2014} serve as reference implementations for blockchain, while Nano (previously known as RaiBlocks)~\cite{Lemahieu2008} is used to represent DAG\footnote{Other DAG approaches are IOTA~\cite{Popov2017} and Byteball~\cite{Churyumov}.}. The listed systems are chosen as representative solutions because of a relatively mature implementation with a notable developer community.  

The rest of the paper is organized as follows: Section~\ref{sec:structure} compares the data structures used to maintain a ledger based on representative implementations. Section~\ref{sec:consensus} explains the algorithm for reaching a consensus within a set of nodes maintaining the ledger. Section~\ref{sec:transaction_confirmation_confidence} examines confidence levels for the inclusion of new entries into the ledger. Section~\ref{sec:ledger_size} analyses the ledger ever growing size due to its append only nature, and explains how ledger size may be reduced over time. Section~\ref{sec:scalability} discusses the scalability issues which DLTs are facing, as one of the main barriers for their global adoption. ~\cref{sec:conclusion} concludes the paper.

\section{Ledger Data Structures}\label{sec:structure}

This section analyzes data structures that are being used to sustain a distributed ledger. Both DAG and blockchain store \textit{transactions} in an open ledger. A ledger has its \textit{state}. Transactions serve as inputs that cause the change to the state, hence DLTs can generally be regarded as \textit{transaction-based state machines}. However, the two approaches use distinct data structures for maintaining the ledger. While blockchain stores transactions in blocks, DAG stores transactions in nodes. The following subsections explain and compares the two data structures.

\subsection{Blockchain}
\begin{figure}[h]
    \centering
    \includegraphics[width=\columnwidth]{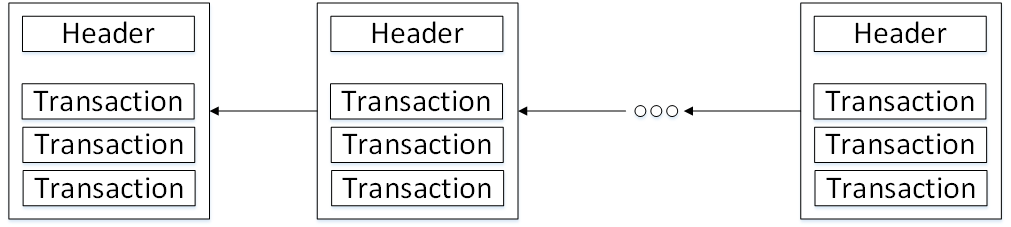}
    \caption{Blockchain as a data structure.}
    \label{fig:blockchain_as_a_data_structure}
\end{figure}

Blockchain consists of ordered units called \textit{blocks}~\cite{DanielDrescher2017}. Blocks contain \textit{headers} and \textit{transactions}, as depicted in Figure~\ref{fig:blockchain_as_a_data_structure}. Each block header, amongst other metadata, contains a reference to its predecessor in the form of the predecessor's hash. The initial state is hard-coded in the first block called the \textit{genesis block}. Unlike other blocks, the genesis block has no predecessor. 

Transactions in Bitcoin and Ethereum are hashed in Merkle Trees \cite{Narayanan2017}. Bitcoin hashes transactions~\cite{Nakamoto2008} in a single tree, while Ethereum uses three different structures to store transactions, receipts and state~\cite{Wood2014}. These structures are reviewed further in Section~\ref{sec:ledger_size_blockchain} in order to explain how to decrease ledger size.

\subsection{Directed Acyclic Graph}

\begin{figure}[h]
    \centering
    \includegraphics[width=0.9\columnwidth]{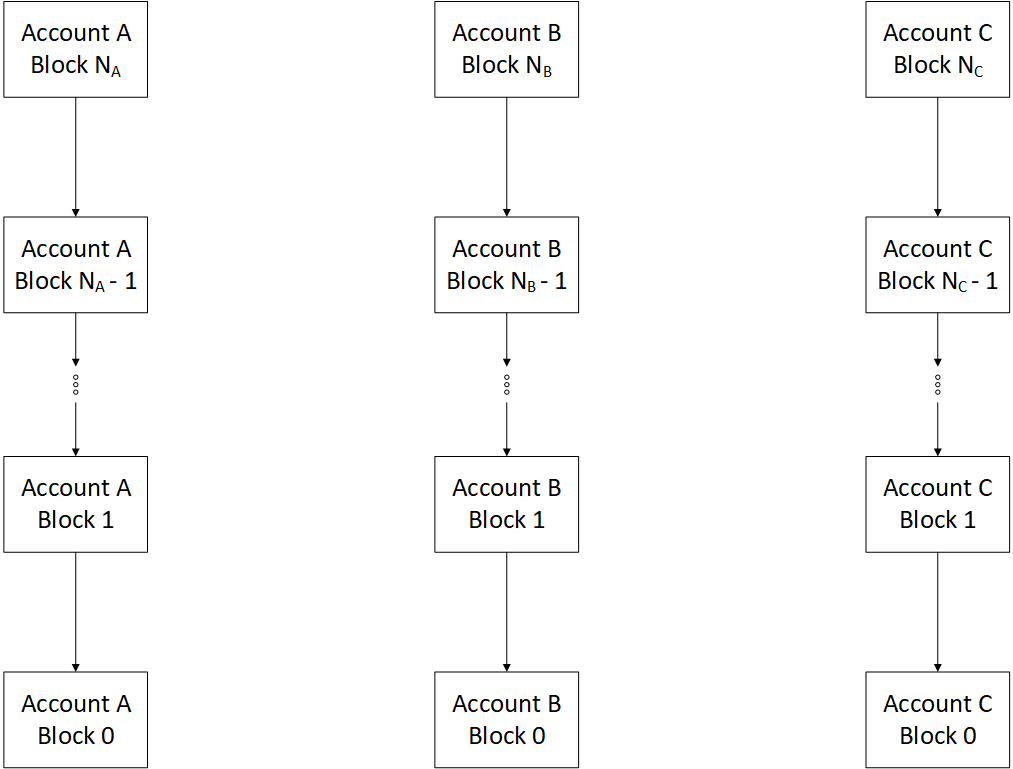}
    \caption{Nano's DAG, the block-lattice.}
    \label{fig:dag_as_a_data_structure}
\end{figure}

In contrast to blocks, a DAG structure stores transactions in \textit{nodes}, where each node holds a single transaction. In Nano, every account is linked to its own account-chain in a structure called the \textit{block-lattice} equivalent to the account’s transaction/balance history. The structure of the block-lattice is displayed in Figure~\ref{fig:dag_as_a_data_structure}. Each account is granted an \textit{account chain}. An account chain can be considered as a dedicated blockchain, just for a single account. Nodes are appended to an account-chain, each node representing a single transaction on the account chain. Similar to the genesis block in blockchain, a DAG holds a \textit{genesis transaction}. The genesis transaction defines the initial state.

\begin{figure}[h]
	\centering
	\includegraphics[width=0.9\columnwidth]{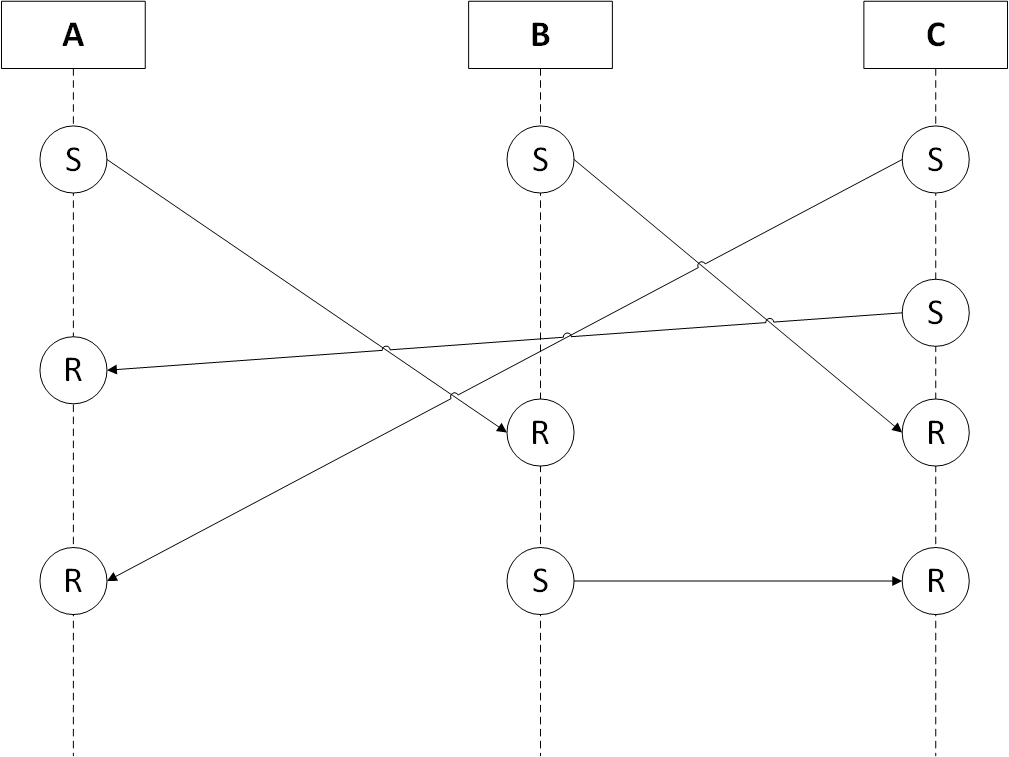}
	\caption{Transaction handling in the block lattice. \textit{S} represents a \textit{send} transaction, \textit{R} represents a \textit{receive} transaction.}
	\label{fig:block_lattice_transactions}
\end{figure}

In Nano, instead of having a single transaction that transfers value, two transactions are needed to fully execute a transfer of value. A sender generates a \textit{send} transaction, while a receiver generates a matching \textit{receive} transaction, as depicted in Figure~\ref{fig:block_lattice_transactions}. When a send transaction is issued, funds are deducted from the balance of the sender's account, and are pending in the network awaiting for the recipient to generate the corresponding receive transaction. While in this state, transactions are deemed \textit{unsettled}. When the receive transaction is generated, the transaction is \textit{settled}. The downside of this approach is that a node has to be online in order to receive a transaction.

\section{Consensus}\label{sec:consensus}

In a \textit{public} and \textit{permissionless} environment where each node can read from the ledger and append to the ledger, blocks or nodes can be malicious and can not be implicitly trusted~\cite{DanielDrescher2017, BitFuryGroup2015}. Bitcoin, Ethereum and Nano are all public and permissionless solutions, and hereafter we discuss consensus mechanisms for such environments.

\subsection{Blockchain}

For an entry to be appended to the ledger, \textit{consensus} about the entry needs to be reached in the network, that is, an agreement must be reached regarding the validity of a new entry that is to be appended to the ledger by all nodes. The assumption is that a supermajority of nodes are honest and reliable.

Algorithms for achieving consensus with arbitrary faults generally require some form of voting among a known set of participants. One method, often referred to as the \textit{Nakamoto consensus}, elects a leader by some form of a lottery~\cite{Sawtooth2018}. The leader then proposes an entry that can be added to the ledger containing a list of previously committed entries. The entries are checked for validity by all other nodes and their consistent ordering is verified. Both Bitcoin~\cite{Franco2014} and Ethereum~\cite{Wood2014} are based on a lottery function called the \textit{Proof of Work} (PoW) (Ethereum has announced to support \textit{Proof of Stake} (PoS) in near future~\cite{Dinh2017}). The elected leader broadcasts the new entry to the rest of the participants who implicitly vote to accept the entry by adding it to their local copy of the ledger, and may propose subsequent transaction entries that build on the ledger~\cite{Sawtooth2018}.

\subsubsection{Proof of Work}
In Proof of Work, the first participant to successfully solve a cryptographic puzzle wins the leader-election lottery. For example, Bitcoin uses partial hash inversion as the cryptographic puzzle function. Partial hash inversion requires that the hash of a block of transactions together with a \textit{nonce} (a free variable in the function) matches a certain pattern. The pattern starts with at least a predefined number of 0 bits~\cite{Franco2014}. The function is difficult to solve  intentionally since to manipulate the ledger, an attacker would need to have the supermajority of the computing power in the network, which makes an attack expensive to perform. Nodes that generate blocks in a Proof of Work driven systems are called \textit{miners} and the process is called \textit{mining}. For the use of their resources, miners are granted \textit{tokens} in the network, as an economic incentive to mine (Ether in Ethereum, Bitcoin in Bitcoin). If there are no miners, no blocks can be mined and there is no transaction throughput. 

\subsubsection{Proof of Stake}

While miners in a PoW driven system commit their computational resources to be elected for block generation, in a PoS driven system users stake their tokens to be able to create blocks. In Ethereum, PoS is implemented in the form of a smart contract named Casper~\cite{Dinh2017}. Validators deposit their stake in the smart contract, which in turn picks the validator allowed to create a block. The more tokens a validator stakes, it has a higher chance to create the next block. If an incorrect block is submitted (e.g., it contains double spending transactions), the validator's stake is \textit{burned}, thus penalizing the validator.

PoS has its advantages over PoW. Firstly, it consumes far less electricity than PoW. For example, based on a recent analysis, Bitcoin mining consumes more electricity in a year that a selected set of 159 countries~\cite{BitcoinMiningCountries2018}). Secondly, attacks on the network are easily penalized relative to PoW. After an attack on a PoW driven network, the attacker still owns the hardware used for the attack. In PoS, burning stake has the same economic effect as dismantling an attackers mining equipment.

\subsection{Directed Acyclic Graph}\label{sec:consensus_dag}

In Nano, there is no need for a leader election since users are obligated to order their own transactions. PoW is still being used, however not for the sake of leader election (since there is none). In the context of Nano, PoW is used as a spam protection measure to prevent over-generation of transactions by a malicious user, similar to Hashcash~\cite{Back2002}. However, a different method for conflict resolution has been introduced, a system of \textit{representatives}. When an account is created, it must choose a representative that can be to changed over time. Representatives vote in order to resolve conflicts. Their votes are \textit{weighted}: a representative's weight is calculated as the sum of all balances for accounts that chose this representative. In the case of a conflict, the wining transaction is the one that gained the most votes with regards to the voters weight~\cite{Lemahieu2008}. For a transaction with no issues, no voting overhead is required.

\section{Confidence of Transaction Confirmation }\label{sec:transaction_confirmation_confidence}

\subsection{Blockchain}\label{sec:transaction_confirmation_confidence_blockchain}

As stated in Section~\ref{sec:consensus}, PoW uses a stochastic process which makes it impossible to know which node will be elected as a leader. Furthermore, even though an entry has been added to the ledger, there is no guarantee that it will remain a valid entry. This seems to be counter intuitive to the inherent feature of distributed ledger, \textit{immutability}. However, it is indeed expected that a ledger may find itself in a temporary state where there are two different histories stored within the ledger. This phenomena is called a \textit{soft fork}~\cite{Sikorski-2017} in blockchain. The issue is eventually resolved by abandoning one version of history over another.

\begin{figure}[h]
	\centering
	\includegraphics[width=\columnwidth]{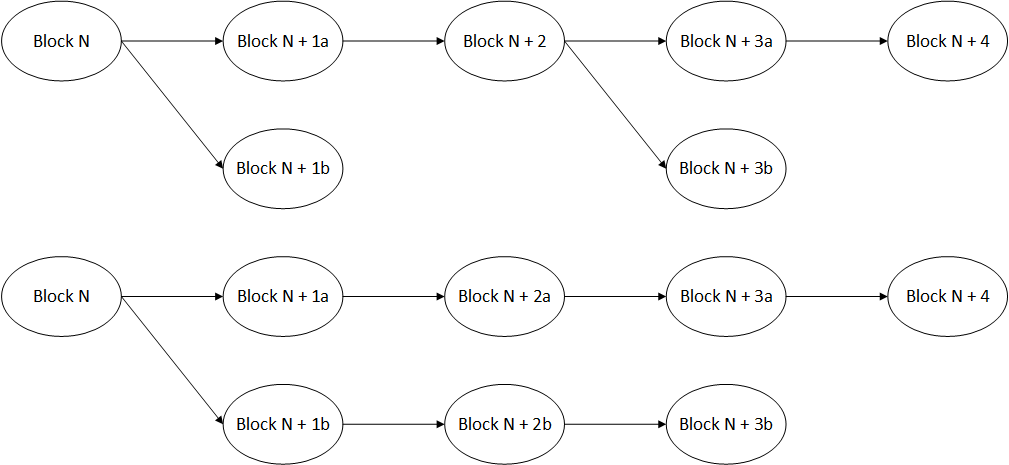}
	\caption{Diagram demonstrating temporary Blockchain forks. The top chain depicts a typical fork, while the bottom chain depicts an atypical fork.}
	\label{fig:blockchain_forks}
\end{figure}

Figure~\ref{fig:blockchain_forks} depicts forks in a blockchain. A soft fork can occur when two different blocks are created at roughly at the same time. Due to network delays, some nodes will receive one block over the other, resulting in a state where two blocks claim the same predecessor. For the time being, nodes continue to build the chain on top of their received blocks, effectively creating~\textit{two chains} possibly containing~\textit{conflicting transactions}. The problem resolves itself when a block is mined that makes one chain longer than the other. The longer chain is adopted, while the shorter one is discarded or~\textit{orphaned}, along with all transactions within it. Orphaned transactions need to be included in a new block.

Since a soft fork can occur at any time, if a block has been appended to the chain, there is no guarantee that it will not get orphaned. As the chain increases in length over the referent block, the probability of the block being discarded decreases. Depending on the implementation, there is a suggested number of blocks that need to be appended above the referent one before it is safe to say that it will remain in the chain with great certainty. The number of appended blocks that guarantee block inclusion with high probability are six for Bitcoin~\cite{Dinh2017} and five to eleven for Ethereum~\cite{Natoli2016}. It is worth mentioning that Ethereum is soon to introduce Casper FFG~\cite{Buterin2017}, a proof of stake based finality system that is supposed to introduce non-reversible checkpoints, guaranteeing block inclusion.

\subsection{Directed Acyclic Graph}\label{sec:transaction_confirmation_confidence_dag}

In Nano, nodes can create transactions at their own discretion at any point in time. However, inconsistencies similar to those in blockchain are still possible. For example, two transactions may claim the same predecessor causing a fork (forks in Nano are only possible as a result of a malicious attack or bad programing) or a transaction may not have been properly broadcasted, causing the network to ignore all subsequent transactions on top of the missing block. When an inconsistency occurs, representatives are called to vote following the procedures explained in Section~\ref{sec:consensus_dag}. 

It is important to note that even though a transaction may be deemed settled, it is only confirmed when it receives a majority vote for the send and receive transactions. Beside voting on conflicts, representatives vote automatically on blocks they have not seen before. A representative that sees a new transaction forwards the transaction with its vote-signature attached if the transaction is valid. This means that the network automatically broadcasts consensus information, while the transaction is making its way through the network. A feature that is supposed to be implemented in the future is \textit{block-cementing} which will prevent transactions from being rolled back after a certain period of time, guaranteeing thus transaction finality~\cite{Lemahieu2008}.

\section{Ledger size}\label{sec:ledger_size}

As every ledger contains all information since its genesis, its size is constantly increasing. With further penetration of the technology, the size will increase even faster. Bitcoin is estimated to be 145,95 GB in size on 02.01.2018~\cite{BitcoinCharts2018}, Ethereum 39.62 GB on 02.01.2018~\cite{EthereumSize2018}. Nano's ledger size is 3.42GB with around $6,700,078$ blocks on 25.02.2018~\cite{NanoLedgerStats2018-1}\cite{NanoLedgerStats2018-2}. In this section we investigate how reference implementations tackle the issue of increasing ledger size.

\subsection{Blockchain}\label{sec:ledger_size_blockchain}

Bitcoin clients offer a pruning mode, allowing users to delete raw block data after the entire ledger has been downloaded and validated, keeping only a small subset of the data. The data is kept in order to be able to relay recent blocks to peers and handle soft forks. The advantage of the method is that disk space is saved. The downside is that other nodes are no longer able to download the entire history of a pruned node~\cite{BitcoinPruningRelease2018}.

Similar to Bitcoin's method of ledger pruning, Ethereum offers a pruning mechanism. Ethereum keeps track of the deltas in the global state maintained by a Merkle state tree. A delta in a global state is the difference between two states of the ledger. Changes made to the state are kept in the ledger in the case of a soft fork, when a state needs to be rolled back, and then updated correctly by the miners on the orphaned branch. However, if one is not interested in past states, the deltas can be discarded without harming the chain integrity. A \textit{fast sync} algorithm has been implemented to tackle this issue. Instead of processing the entire blockchain one link at a time and replaying all transactions that ever happened in history, fast syncing downloads the transaction receipts along the blocks, and pulls an entire recent state~\cite{GethFast2018}. After downloading a state which is recent enough (head of the chain - 1024 blocks, also called the \textit{pivot point}), the process is paused for state sync where the Merkle state tree is downloaded from the pivot point. For every account found in the tree, it's contract code and internal storage state tree is retrieved. From the pivot point onward, all blocks are downloaded and the node continues its usual operation. The result of the mechanism is a database pruned of the state deltas.

\subsection{Directed Acyclic Graph}

Nano distinguishes between three types of nodes: \textit{historical} which keep record of all transactions, \textit{current} which keep only the head of account-chains, and \textit{light} that do not hold any ledger data and only observe or create new transactions (in the current implementation, all nodes are historical nodes). 

In order to reduce the ledger's size, Nano plans to implement \textit{pruning}. Since the accounts keep record of account balances instead of unspent transaction inputs, all other historical data can be discarded to decrease ledger size. This feature is yet to be implemented in 2018~\cite{RaiBlocksRoadmap2018}.

\section{Scalability}\label{sec:scalability}

One of the most relevant issues hindering global scale DLT adoption is it's scalability. At 05.01.2018. there were around $186,951$ pending transactions in the Bitcoin network~\cite{BitcoinCharts2018} and around $22,473$ pending in the Ethereum network~\cite{Etherscan2018}. This section explains how the two technologies handle incoming transactions in terms of scalability.

\subsection{Blockchain}

In order for a transaction to be included in a block (included being different from confirmed, see Section~\ref{sec:transaction_confirmation_confidence_blockchain}), a block must be created. A block is created every time when a PoW puzzle is solved, a thus transaction rate is limited by the periodicity at which blocks are created and also by the block size. When increasing the number of nodes in the system, the frequency of block creation does not increase significantly due to the fact that the PoW puzzle difficulty is dynamic so that the block generation time converges to a fixed value.  

In Bitcoin, a block is mined roughly every 10 minutes with a maximum block size of 1 MB, thereby limiting the Bitcoin transaction rate to between 3 and 7 transactions per second, depending on the size of individual transactions on the blockchain~\cite{Hari2016}~\cite{Cromanetal2016}.

In Ethereum, a block is mined roughly every 15 seconds~\cite{EthereumScailing2018} with a dynamic block size not measured in bytes but rather in \textit{gas}. Gas is the unit used to measure the fees required for a particular computation~\cite{Wood2014}. In the context of Ethereum block size, a measure called \textit{gas limit} defines the maximum amount of gas all transactions in the whole block combined are allowed to consume. In contrast to Bitcoin, this value is dynamic and will adapt to network conditions. This enables Ethereum's transaction rate to be roughly between 7 to 15 transactions per second~\cite{EthereumSharding2018}. The transition to PoS should decrease Ethereums block generation time to 4 seconds or lower~\cite{PoSTransactionSpeed2018}. However, this is still a rather limited block generation rate.

Since Bitcoin and Ethereum are used for payments, it is interesting to compare them with already existing payment solutions, such as \textit{Visa} which is able to process 56,000 transactions per second~\cite{Visa2015}. Note also that Ethereum has a significant benefit compared to Bitcoin since it supports \textit{smart contracts}~\cite{Wood2014}, which expands its potential to become a platform rather than only a cryptocurrency.

A potential approach to improve scalability is to increase the block size (be it in megabytes or in gas limit). Increasing the block size also increases the maximum amount of transactions that fit into a block, effectively increasing transaction rate. However, the block size increase would eventually lead to centralization due to the fact that consumer hardware would become unable to process blocks leading to the network relying on supercomputers~\cite{EthereumSharding2018}. One of such efforts is Segwit2x~\cite{Segwit2x2018} in Bitcoin which, among other features, tries to increase the block size to 2MB.

Another approach is to create \textit{channels}, scaling the transaction capacity. One such implementation is the Raiden Network~\cite{Raiden2018} on top of Ethereum or the Lightning Network~\cite{Payments2016} on top of Bitcoin. The solution revolves around creating an off chain channel to which a prepaid amount is locked in for the lifetime of the channel. The involved parties are able to run micro transactions at high volume and speed, avoiding the transaction cap of the network. Any party may choose to leave the channel, after which the final account balances are recorded on chain and the channel is closed.

Another attempt to increase scaling in Ethereum is Plasma~\cite{Poon2017}. The framework creates a nested blockchain structure by the use of smart contracts with a root chain being the Ethereum main chain. Constraints and consensus mechanisms are defined by a smart contract and based on PoS. Only Merkle roots created in the sidechains are periodically broadcasted to the main network during non-faulty states allowing scalable transactions. 
For faulty states, stakeholders need to display proof of fraud and the Byzantine node gets penalized. An example network being written for the Plasma framework is OmiseGO~\cite{Poon2017b}.

A more complex approach to further improve scalability is \textit{sharding}. Sharding splits the network in K partitions, no longer forcing all nodes in the network to process all incoming transactions. Every shard $k \in K$, in it's simplest form, has it's own transaction history and the effects of a transition in shard $k$ would effect only the state of $k$. In a more complex scenario, cross shard communication is available, meaning that for $k,m \in K, k \neq m$ a transaction from $k$ can trigger an event in $m$~\cite{EthereumSharding2018}. The downside of this approach is that developers would need to be aware that they are programming in a cross shard environment. The Ethereum foundation is attempting to make cross shard communication \textit{transparent} for developers~\cite{EthereumSharding2018}, which will in turn further increase the complexity of the protocol.

\subsection{Directed Acyclic Graph}\label{sec:scalability_dag}

Opposed to blockchain technology where dedicated validators must exist in order to generate and order blocks, a user in Nano must sort his/her own transactions. 
This approach vastly differs from the way transactions are executed on blockchain systems. Namely, instead of having validating nodes charged with transaction ordering, transaction ordering is done asynchronously by the account owner being in charge of transaction ordering. This approach greatly influences scalability. The consequence of this design decision is that there is no inherent cap in the transaction throughput in the protocol itself. However, peak throughput on a test reached on the main network was $306$ Transactions Per Second (TPS) with an average of $105.75$ TPS~\cite{RaiBlocksStress2018-2}. The limit is currently determined by the quality of consumer grade hardware and network conditions.

\section{Conclusion}\label{sec:conclusion}

When comparing DAG and blockchain based ledgers, one can conclude that DAG based ledgers store transactions as edges in an directed acyclic graph while blockchains bundle transactions in blocks and append blocks one after another. Blockchain technology determines the global truth by choosing a single branch that holds all the transactions.
Global truth and transaction ordering in public and permissionless blockchains is generally done by some sort of leader election, either using PoW or PoS. Leaders are elected stochastically and the global truth is found in the longest chain, while a shorter one is abandoned. Nano's DAG abandons leader election and delegates transaction ordering to users and their representatives to resolve conflicts.

Due to the fact that a branch in a blockchain may become orphaned, just the fact that a transaction is included in a block doesn't mean that it will remain in the ledger version containing the global truth. For that reason, it is recommended to wait for some number of blocks to be appended above the referent one before concluding that it is confirmed. In Nano's DAG, a transaction is confirmed when there is a majority of votes cast in favor of a transaction by the representatives. 

Increased ledger size is a significant problem for all DLTs and this issue is tackled by ledger pruning. The entire history is federated to historical nodes while other nodes only maintain a subset of historical data. Generally, a tradeoff between disk space usage and historical data accessibility is being made.

A scalable DLT can be defined as a system where every node does not need to process every transaction, and thus existing DAG or blockchain implementations do not guarantee scalability per se. This paper describes how existing blockchain and DAG implementations try to achieve scalability: Blockchain solutions propose the following approaches: increased block size, support of off-chain channels, hierarchical chains and ultimately sharding. DAGs can impove scalability by coupling network usage and transaction verification, meaning that a user must handle his/hers own transactions in order to use the network. Even though theoretically uncapped protocols for achieving global consensus exist (e.g. Nano's consensus protocol is theoretically uncapped while the Bitcoin network creates a block every 10 minutes), one must take into account real world limitations, e.g., network conditions and processing power.

%Može li se zapisati kakav generalni zaključak, naglasiti doprinos članka i navesti future work?

% conference papers do not normally have an appendix

% use section* for acknowledgment
\section*{Acknowledgment}

The authors would like to thank the Nano community for their cooperation and responsiveness.

% trigger a \newpage just before the given reference
% number - used to balance the columns on the last page
% adjust value as needed - may need to be readjusted if
% the document is modified later
%\IEEEtriggeratref{8}
% The "triggered" command can be changed if desired:
%\IEEEtriggercmd{\enlargethispage{-5in}}

% references section

% can use a bibliography generated by BibTeX as a .bbl file
% BibTeX documentation can be easily obtained at:
% http://mirror.ctan.org/biblio/bibtex/contrib/doc/
% The IEEEtran BibTeX style support page is at:
% http://www.michaelshell.org/tex/ieeetran/bibtex/
%\bibliographystyle{IEEEtran}
% argument is your BibTeX string definitions and bibliography database(s)
%\bibliography{IEEEabrv,../bib/paper}
%
% <OR> manually copy in the resultant .bbl file
% set second argument of \begin to the number of references
% (used to reserve space for the reference number labels box)
\bibliographystyle{IEEEtran}
\bibliography{Blockchain,RandomBlockchain,Whitepapers}

% that's all folks
\end{document}